\documentclass{sty/IEEEtran}
\usepackage{cite}
\usepackage{amsmath,amssymb,amsfonts}
\usepackage{algorithmic}
\usepackage{graphicx}
\usepackage{textcomp}
\def\BibTeX{{\rm B\kern-.05em{\sc i\kern-.025em b}\kern-.08em
    T\kern-.1667em\lower.7ex\hbox{E}\kern-.125emX}}
\ifCLASSOPTIONcompsoc
\usepackage[caption=false, font=normalsize, labelfont=sf, textfont=sf]{subfig}
\else
\usepackage[caption=false, font=footnotesize]{subfig}
\fi

\usepackage{cite}
\usepackage{threeparttable}
\usepackage{multirow}
\usepackage{times}
\usepackage{float}
\usepackage{enumerate}
\usepackage[none]{hyphenat}
\usepackage[normalem]{ulem}
\usepackage{color}
\usepackage{xcolor}

\usepackage{tikz}
\usepackage{pgfplots}
\pgfplotsset{compat=newest}
\usetikzlibrary{positioning}

\usepackage{algorithm}
\usepackage{algorithmic}

\newcommand{\qed}{\nobreak \ifvmode \relax \else
	\ifdim\lastskip<1.5em \hskip-\lastskip
	\hskip22em plus0em minus0.5em \fi \nobreak
	\vrule height0.4em width0.3em depth0.25em\fi}

\newlength\fheight
\newlength\fwidth
\definecolor{green}{rgb}{0.0, 0.5, 0.0}

\newcommand{\Fig}[1]{Fig.\,{\ref{#1}}}

\begin{document}
\title{{\fontsize{24}{26}\selectfont{Communication\rule{29.9pc}{0.5pt}}}\break\fontsize{16}{18}\selectfont
Three-Dimensional Fully Metallic Dual Polarization Frequency Selective Surface Design Using Coupled-Resonator Circuit Information}

\author{Ignacio Parellada-Serrano, Mario P\'{e}rez-Escribano, Carlos Molero, Pablo Padilla and Valent\'{i}n de la Rubia
    \thanks{I. Parellada-Serrano, C. Molero, and P. Padilla are with the Department of Signal Theory, Telematics and Communications, Research Centre for Information and Communication Technologies (CITIC-UGR), University of Granada, Granada, Spain (e-mails: parellada@ugr.es; cmoleroj@ugr.es; pablopadilla@ugr.es).

    M. P\'{e}rez-Escribano is with the Telecommunication Research Institute (TELMA), Universidad de M\'{a}laga, E.T.S. Ingenier\'{i}a de Telecomunicaci\'{o}n, 29010 Málaga, Spain (e-mail: mpe@ic.uma.es).

    V. de la Rubia is with the Departamento de Matem\'{a}tica Aplicada a las {TIC}, ETSI de Telecomunicaci\'{o}n, Universidad Polit\'{e}cnica de Madrid, 28040 Madrid, Spain (e-mail: valentin.delarubia@upm.es).

    Manuscript received XX/XX/XXXX; revised XX/XX/XXXX; accepted XX/XX/XXXX. This work has been supported by grant PID2020-112545RB-C54 funded by MCIN/AEI/10.13039/501100011033 and by the European Union NextGenerationEU/PRTR. It has also been supported by grants PDC2022-133900-I00, TED2021-129938B-I00 and TED2021-131699B-I00, and by Ministerio de Universidades and the European Union NextGenerationEU, under Programa Margarita Salas.}}
    

\maketitle
\begin{abstract}


This work employs a new approach to analyze coupled-resonator circuits to design and manufacture a fully metallic dual polarization frequency selective surface (FSS). The proposed filtering structure is composed of a series of unit cells with resonators \emph{fundamentally} coupled along the \emph{z}-direction and then repeated periodically in the \emph{xy}-plane. The fully metallic cascaded unit cell is rigorously analyzed within an infinite periodic environment as a coupled-resonator electromagnetic (EM) circuit. The convenient design of the EM resonators makes it possible to push the evanescent EM field through the metallic structure in the desired frequency band for both polarizations. An FSS prototype is manufactured and measured, and good agreement is found between the simulation results and the final prototype.
\end{abstract}
\markboth{IEEE Transactions on Antennas and Propagation}{Parellada et. al.: Three-dimensional Fully Metallic Dual Polarization Frequency Selective Surface Design Using Coupled-Resonator Circuit Information}
\section{Introduction} \label{sec:Introduction}
Evanescent filters emerged as technological solutions to overcome drawbacks associated with propagating filters \cite{Lecouve2000}. Evanescent filters significantly improve insertion-loss levels and provide a flatter passband and sharp rejection responses (good selectivity) \cite{Reiter1977} thanks to the inherent high Q-factor. It otherwise invokes narrow-band transmission, which is very useful for the output stages in data transmitters \cite{Subramanyam2015}. In addition, evanescent filters outperform in a compact size and reduced weight \cite{Labay1992}, enabling a suitable integration in communication systems \cite{Craven1969}.

Pioneers on this topic date back to the fifties. To the authors' best knowledge, the first complete publication about filters based on evanescent networks was realized by S. B. Cohn in 1957 \cite{Cohn1957}, who explored lumped-elements topologies. Later on, Prof. Craven and his team further studied these systems \cite{Craven1966, Craven1969, Craven1971}, both from the theoretical and experimental point of view. Rectangular waveguides were established as the prominent technology for evanescent filters due to their simplicity to operate in the cutoff (evanescent) region \cite{Snyder1977, Schunemann1977}. The development of this technology has continued until the present century, benefiting from the permanent improvement of fabrication techniques and commercial electromagnetic solvers. Special attention deserves the in-line configurations \cite{Amari2003}, consisting on waveguides loaded with periodically spaced pin-loads \cite{Kirilenko1999}, ridges \cite{RuizCruz2005, RuizCruz2006}, dielectric-mushrooms \cite{Singhal2022}, exotic ridges \cite{Polo2022}, non-resonating modes \cite{Bastioli2021,Bastioli2023}, and frequency-variant couplings \cite{Macchiarella2021, Mrozowski2023}, among others.

Filters based on FSSs are modern solutions proposed to operate in certain scenarios, such as specific spacial and military environments demanding applications for RCS reduction in aircraft or antenna radomes \cite{Liu2018, Martini2006, Zhou2012}. FSSs exhibit more flexibility in covering large areas or adapting to curved surfaces \cite{Fernandez2021}. They are otherwise generally based on propagating systems \cite{Behdad2008}, lacking good sensibility and increasing the risk of non-desired interferences. Full-metal 3D designs, as those in \cite{Molero2019}, arise as promising solutions. Full-metal architectures are more robust to extreme thermal and environmental conditions, being promising candidates for spatial missions \cite{Chahat2018}. In addition, full-metal cells have high Q-factors and may operate in cutoff regime \cite{Molero2020, Velasco2021}, satisfying low-weight, small-size performance and flat-band and sharp-rejection responses.  

The structure proposed in this paper is a fully metallic FSS, as shown in Fig.~\ref{whole_cell}. It is formed by periodic arrangements of square waveguides with dog-bone resonators perforated along the walls. We started from previous FSS design topologies, such as \cite{Molero2020}. This FSS structure has never been employed for filtering purposes. Furthermore, addressing more than three cascaded resonators in the design becomes a complicated task since controlling the couplings among resonators turns into a real challenge, never mind taking into account both polarizations, where even more resonances need to be handled. This paper proposes a dual polarization FSS design with a wideband performance by increasing the number of resonators to 7 along each polarization. As a result, 14 resonances are present within the FSS unit cell. Due to symmetry considerations, the design of a 7$^\text{th}$ order coupled-resonator circuit for each polarization is enough to account for the dual polarization behavior rigorously. A recent electromagnetic coupling matrix technique \cite{delaRubia2022EMCouplingMatrix} is used for both, approve a tentative initial design, and guide in the final full-wave optimization loop to tune the frequency response of the FSS straightforwardly.


\section{Electromagnetic Coupling Matrix} \label{sec:EMCouplingMatrix}

\begin{figure}[t!]
	\centering
	\subfloat[]{\includegraphics[width=0.35\columnwidth]{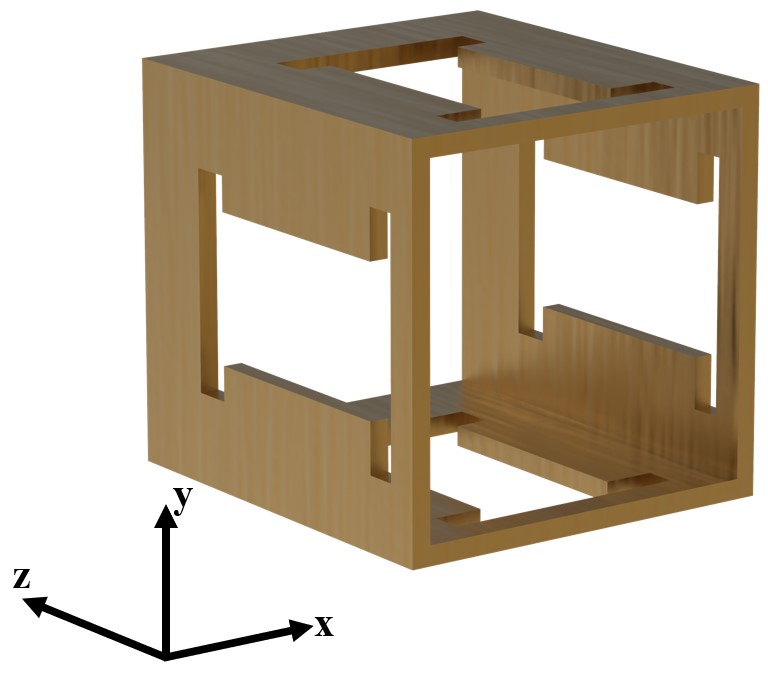} \label{unitcell}}
	\hspace{0mm}
	\subfloat[]{\includegraphics[width=0.35\columnwidth]{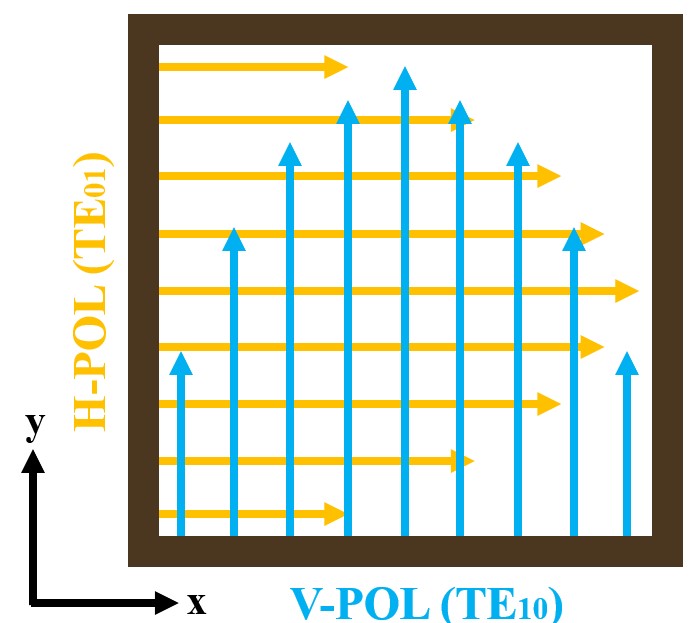} \label{doblepol}}
	\hspace{0mm}
	\vspace{3mm}
	\subfloat[]{\includegraphics[width=0.5\columnwidth]{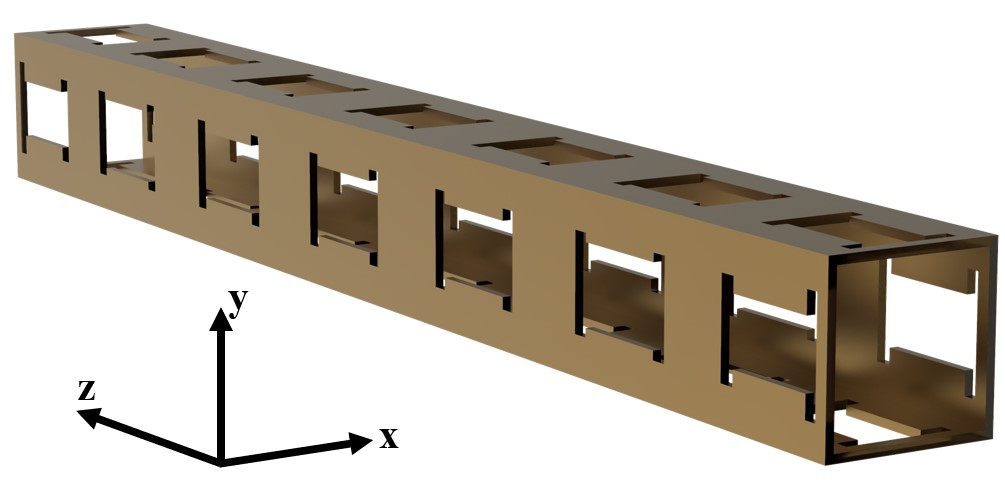} \label{CascadeFilterUnitCell}}
	\caption{\label{whole_cell} 3D unit cell: (a) unit cell model, (b) independent polarization scheme, and (c) cascade filter unit cell.} 
	\label{cell}
\end{figure}

Due to the 90-degree symmetry along the \emph{z}-direction (EM wave propagation direction) taken into account in the unit cell detailed in Fig.~\ref{cell}, this structure can be rigorously analyzed for one \emph{single} polarization, and thus obtaining the response for the orthogonal polarization in the same analysis. This is actually the rationale behind the 90-degree symmetry along the \emph{z}-direction in the unit cell shown in Fig.~\ref{CascadeFilterUnitCell}. Furthermore, under TE plane wave illumination, this periodic unit cell can be effectively analyzed employing \emph{simple} PMC and PEC boundary conditions on the corresponding opposite sidewalls, thus dropping the requirement for periodic boundary conditions. 

Under this scenario, we solve time-harmonic Maxwell's equations in the analysis domain $\Omega \subset \mathbb{R}^3$ (which contains the unit cell as shown in Fig.~\ref{cell}) to obtain the electromagnetic field as a function of frequency (although we prefer wavenumber notation, $k$) in $\Omega$. We carry out this full-wave analysis by means of the finite element method (FEM) and the reduced-basis method, where no approximation is taken into account. \texttt{Gmsh} is used to mesh the analysis domain \cite{GeuR09}. As a result, a reliable representation of the electric field in $\Omega$ for the band of analysis $\mathcal{B}=[k_\text{min},k_\text{max}]$, is obtained, viz.

\begin{equation}
\label{eq:Sec-EMCouplingMatrix-SolutionFrequencyDependency}
\mathbf{E}(k) = j k \eta_0 \sum \limits_{k_n^2 \in \mathcal{B}_2} \frac{A_n }{k_n^2 - k^2 } \mathbf{e}_n + \sum \limits_{n = 1}^{N} \beta_n(k) \mathbf{E}(\kappa_n). \\
\end{equation}
$\mathcal{B}_2$ stands for $[k^2_\text{min},k^2_\text{max}]$. $\eta_0$ is the intrinsic impedance in the vacuum. $k_n$ and $\mathbf{e}_n$ stand for the eigenresonances and corresponding eigenmodes of the FSS unit cell. Coefficients $A_n$ and $\beta_n(k)$ are conveniently determined in the full-wave analysis by the reduced-basis method. We refer the interested reader to \cite{delaRubia2018CRBM} for all the details.

This electric field solution \eqref{eq:Sec-EMCouplingMatrix-SolutionFrequencyDependency} allows us to find the impedance matrix transfer function $\mathbf{Z}(k)$ with ease. Thus,
\begin{equation}
\label{eq:Sec-EMCouplingMatrix-ImpendanceMatrix}
\begin{aligned}
\begin{pmatrix}
v_1 \\
\vdots \\
v_M 
\end{pmatrix} &= j k \eta_0 \sum \limits_{k_n^2 \in \mathcal{B}_2} \frac{ \begin{pmatrix} c_{n1} \\ \vdots \\ c_{nM} \end{pmatrix}   \begin{pmatrix} c_{n1} & \cdots & c_{nM} \end{pmatrix} }{k_n^2 - k^2 } \begin{pmatrix}
i_1 \\
\vdots \\
i_M
\end{pmatrix} \\
&+ \mathbf{Z}_\text{out-of-band}(k) \begin{pmatrix}
i_1 \\
\vdots \\
i_M
\end{pmatrix} \\
\mathbf{v} &= \mathbf{Z}(k) \mathbf{i} =  (\mathbf{Z}_\text{in-band}(k) + \mathbf{Z}_\text{out-of-band}(k) )\mathbf{i} \\
           &= \mathbf{v}_\text{in-band}+\mathbf{v}_\text{out-band} =  \mathbf{Z}_\text{in-band}(k) \mathbf{i} + \mathbf{Z}_\text{out-of-band}(k) \mathbf{i} \text{.}
\end{aligned}
\end{equation}
We have deliberately split the EM contributions into in-band and out-of-band, namely, $\mathbf{Z}_\text{in-band}$ and $\mathbf{Z}_\text{out-of-band}$. As a result, only in-band eigenmodes are taken into account in $\mathbf{Z}_\text{in-band}$, and the remaining EM contributions are left to $\mathbf{Z}_\text{out-of-band}$. $\mathbf{v}$ is also decomposed into these two contributions, namely, $\mathbf{v}_\text{in-band}$ and $\mathbf{v}_\text{out-band}$, respectively. The poles $k_n$ in $\mathbf{Z}_\text{in-band}$ have rank-1 matrix residues, cf. \eqref{eq:Sec-EMCouplingMatrix-ImpendanceMatrix}. This resembles a Foster impedance representation in $\mathbf{Z}_\text{in-band}$ \cite{felsen2009electromagnetic}. This property allows us to find a more insightful state-space dynamical system matrix representation for $\mathbf{Z}_\text{in-band}$, viz.
\begin{subequations}
	\label{eq:Sec-EMCouplingMatrix-DynamicalSystem}
	\begin{align}
	\label{eq:Sec-EMCouplingMatrix-DynamicalSystem1}
	\begin{pmatrix}
	\mathbf{0} & \mathbf{C} \\
	\mathbf{C}^T & \mathbf{A}(k) \\
	\end{pmatrix}
	\begin{pmatrix}
	\mathbf{i} \\
	\mathbf{E} \\
	\end{pmatrix}
	&=
	\begin{pmatrix}
	\frac{\mathbf{v}_\text{in-band}}{-j k \eta_0} \\
	\mathbf{0} \\
	\end{pmatrix}
	\\
	\label{eq:Sec-EMCouplingMatrix-DynamicalSystem2}
	\mathbf{v}_\text{in-band} = j k \eta_0 \mathbf{C} \mathbf{A}^{-1}(k) \mathbf{C}^T \mathbf{i} &= \mathbf{Z}_\text{in-band}(k) \mathbf{i}\text{.}
	\end{align}
\end{subequations}
$\mathbf{A}(k)$ is a diagonal matrix with entries $k_n^2-k^2$, namely, $\mathbf{A}(k)=\mathbf{K}-k^2\mathbf{Id}$, $\mathbf{K}=\text{diag}\{k_n^2 \in \mathcal{B}_2\}$, the state space $\mathbf{E}$ stands for the electric field in the in-band eigenmode basis $\{\mathbf{e}_n, k_n^2 \in \mathcal{B}_2 \}$, and $\mathbf{C}$ matrix entries $C_{pn}$ ($C_{pn}=c_{np}$), stand for the coupling coefficients from ports to each in-band state, i.e., to each eigenmode found in the band of analysis $\mathcal{B}$. As a result, the matrix
\begin{equation}
\label{eq:Sec-EMCouplingMatrix-CouplingMatrix}
\begin{pmatrix}
\mathbf{0} & \mathbf{C} \\
\mathbf{C}^T & \mathbf{K} \\
\end{pmatrix}
\end{equation}
gives rise to an electromagnetic coupling matrix description in the transversal topology of the FSS unit cell in the band of interest $\mathcal{B}$. Further manipulations can be carried out to get the electromagnetic coupling coefficients among resonators. See \cite{delaRubia2022EMCouplingMatrix,delaRubia2022FullWaveCouplingMatrix} for further details.

Summing up, every time a full-wave analysis is carried out within the analysis domain $\Omega$, we get valuable design information \emph{for free} (no additional computations have to be carried out) by using this electromagnetic coupling matrix approach. This design information guides us in the full-wave optimization loop to tune, \emph{in few iterations}, the target EM frequency response of the infinite FSS. We will get back to this point in the next Section. 

%
\section{Baseline Unit Cell} \label{sec:BaselineUnitCell}
The proposed unit cell in this study adopts a fully metallic three-dimensional geometry and is periodically placed on the \emph{xy}-plane. This unit cell is illustrated in \Fig{unitcell} and is influenced by the structure in \cite{Molero2020}. It is the fundamental building block for defining the filtering structure, determining its order, and specifying its properties. To prevent internal propagation, the dimensions of the cell are carefully chosen. The transmission response is then regulated through dog-bone-shaped resonators inserted along the walls. Despite its reactive nature, the $\text{TE}_{10}$-mode (or the corresponding $\text{TE}_{01}$-mode), is excited by the incident polarization, which can be either vertical or horizontal, along $\emph{y}$ or $\emph{x}$, respectively (see Fig.~\ref{doblepol}). The filter order corresponds to the number of resonators along $\emph{z}$-direction for each polarization, which is the direction of propagation.

Including resonators along the propagation direction imparts a three-dimensional nature to the overall structure. This design approach leverages the additional dimension to enhance the performance, granting an extra degree of freedom. This design scheme enables reasonable independent manipulation of each polarization. The horizontal polarization is fundamentally influenced by resonators positioned along the \emph{x}-direction, while the vertical polarization is affected fundamentally by resonators placed along the \emph{y}-direction. Inside the cell, vertical polarization excites the $\text{TE}_{10}$-mode whereas the horizontal one excites the $\text{TE}_{01}$ one as illustrated in Fig.~\ref{doblepol}. Employing exclusively metallic materials in the filtering structure eliminates the inherent losses associated with dielectrics, resulting in improved efficiency. 

\begin{figure}[tbp]
    \centering
    \subfloat[]{\includegraphics[width=0.9\columnwidth]{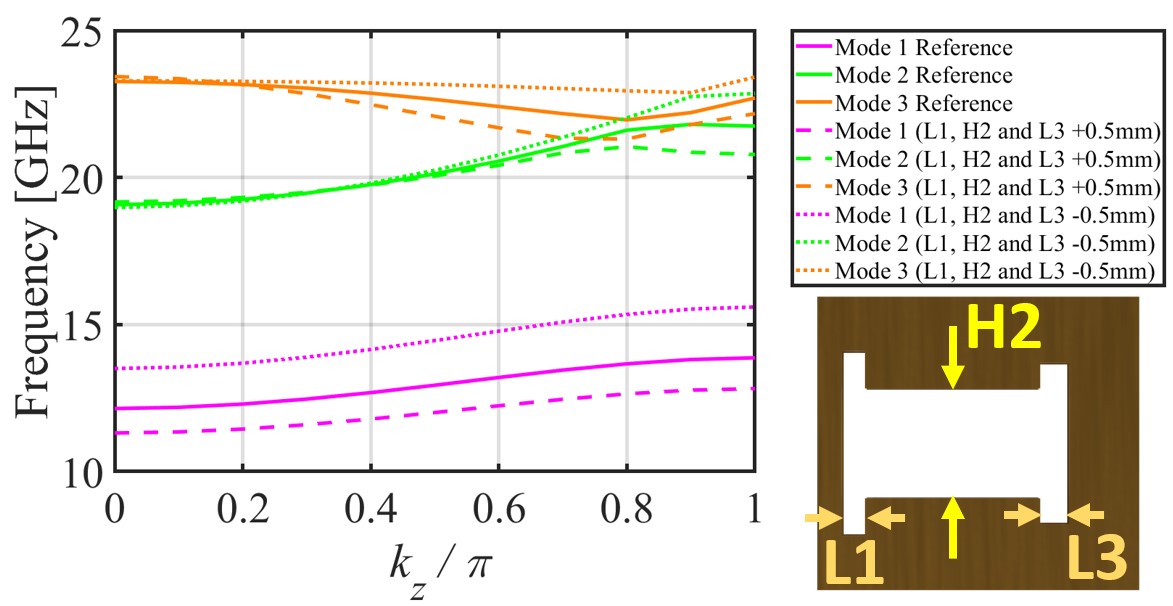}\label{DispersionCelda_Espesor}}
    \hspace{0mm}
    \subfloat[]{\includegraphics[width=0.9\columnwidth]{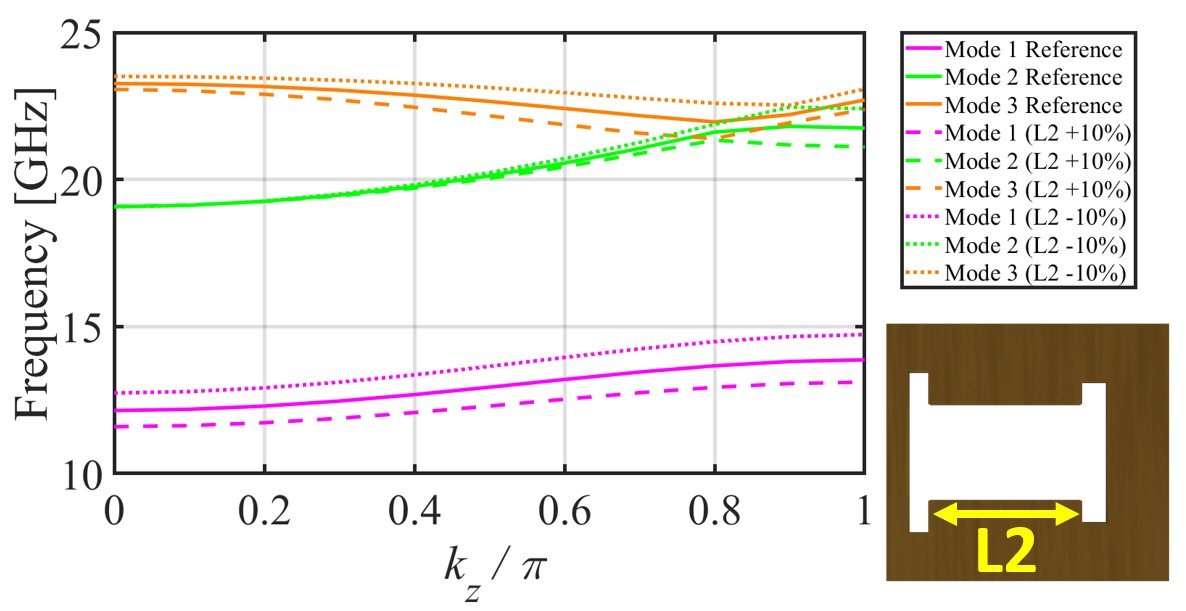}\label{DispersionCelda_Centro}}
    \hspace{0mm}
    \subfloat[]{\includegraphics[width=0.9\columnwidth]{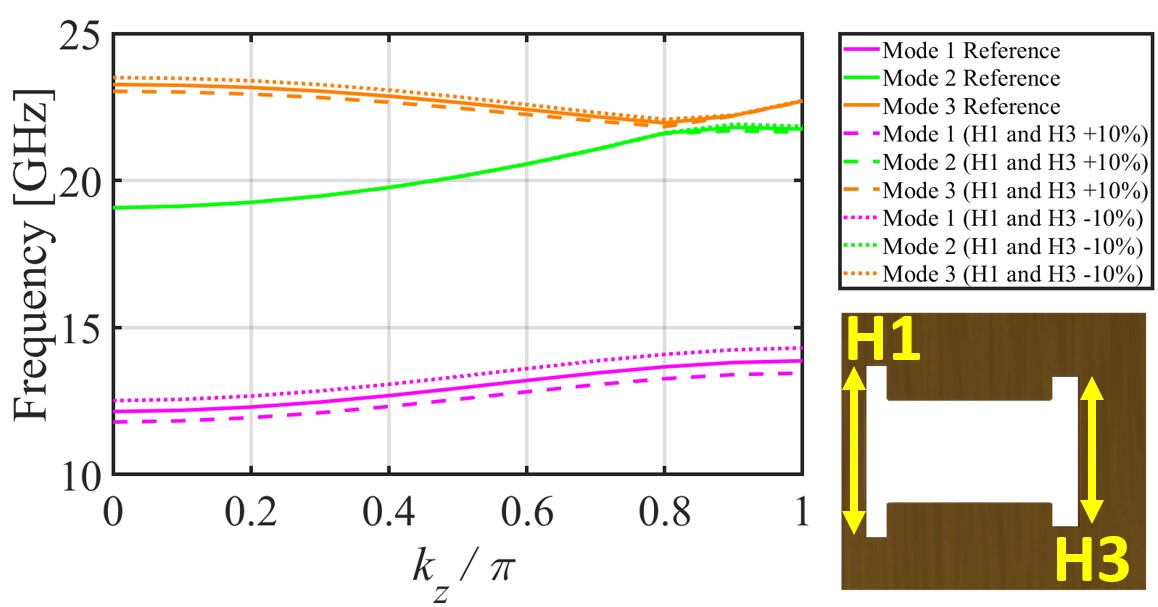}\label{DispersionCelda_Patas}}
    \caption{Dispersion diagram of baseline unit cell and variations, (a) modifying the size of the resonator, (b) modifying the central length of the resonator, and (c) modifying the vertical heights of the resonator.}
    \label{DispersionCelda}
\end{figure}

In Fig.~\ref{DispersionCelda}, a preliminary analysis of the unit cell is conducted, assuming periodicity along the direction of propagation \emph{z}. The dispersion diagram depicts various configurations of the cell, showcasing the changes in its behavior as the dimensions of the resonators are altered. Fig. \ref{DispersionCelda_Espesor} illustrates the effect of modifying the resonator's size within a range of $\pm0.5$~mm, while Figs. \ref{DispersionCelda_Centro} and \ref{DispersionCelda_Patas} demonstrate adjustments to the indicated dimensions within 10$\%$ of their original values. The results presented in Fig. \ref{DispersionCelda} reveal that the reference unit cell establishes a passband for the first and higher-order modes, with a distinct stopband between the first and second modes. The presence of wide stopbands is expected due to the opacity of the cell. However, when the resonators approach a resonance, the opaque character diminishes, opening a passband for the EM field. By varying the parameters, it is possible to manipulate the behavior and frequency range of the first mode. At the same time, more significant changes to the resonator shape allow for modifications in the higher-order modes.

%

After modeling and parameterizing the baseline unit cell, the subsequent step involves its utilization for constructing the cascade filter. To achieve the desired filtering properties in a periodic grid, multiple instances of these unit cells are gathered together, resulting in the formation of our 7$^\text{th}$ order unit cell (cascaded along the \emph{z}-direction) for each polarization. This study concentrates explicitly on a filter created by concatenating 7 unit cells, as depicted in Fig. \ref{CascadeFilterUnitCell}.
 
As it is well-known, increasing the number of concatenated resonators leads to more resonances. Consequently, this makes it possible to keep a low level of $S_{11}$ (below -20~dB in our case) as a design objective in a wide band. Moreover, the transition between the passband and stopband exhibits a more pronounced and abrupt response.

\begin{figure}[tbp]
	\centering
	\includegraphics[width=1\columnwidth]{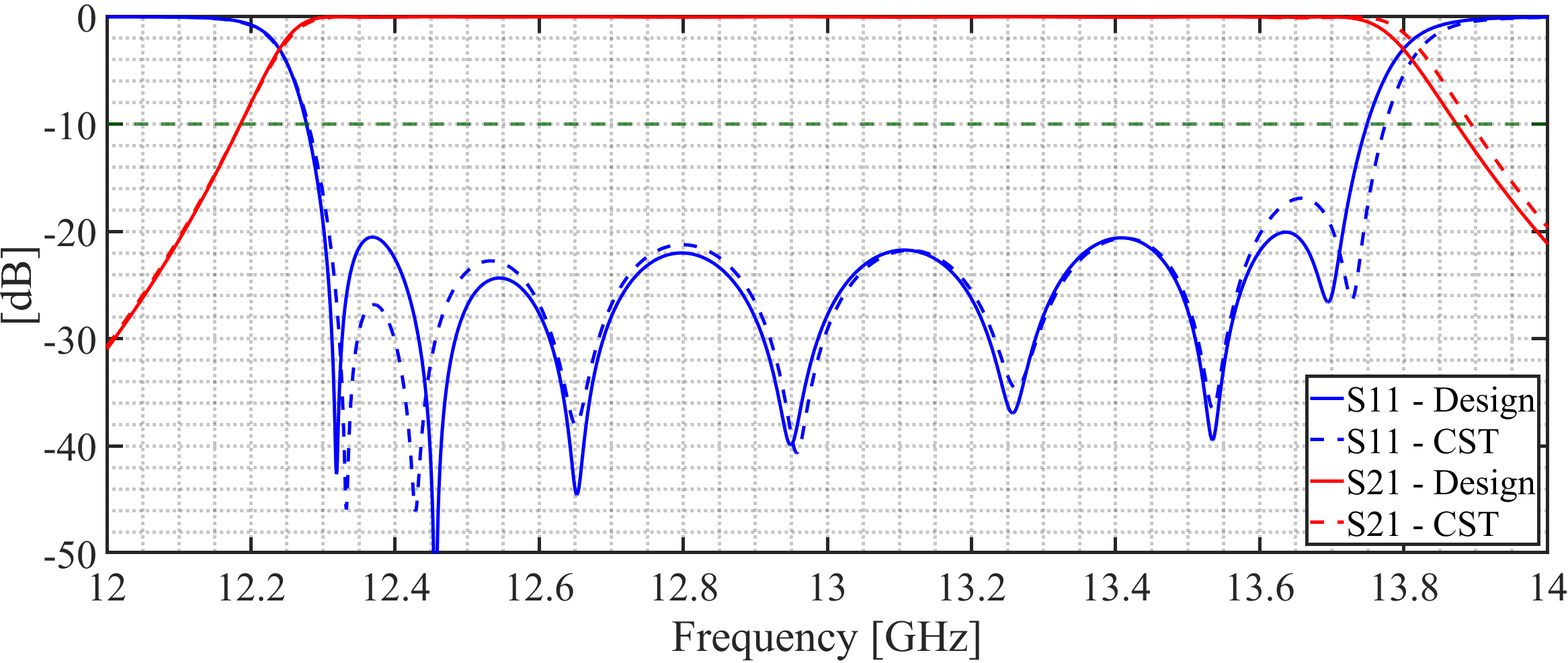}
	\caption{Infinite FSS response: $S$-parameter comparison between design and CST simulation.}
	\label{DesignInfiniteFSS}
\end{figure}

The design process primarily involves determining the dimensions of each resonator to achieve the desired couplings and resonances. In this study, we focus on the design of a passband filter centered at 13~GHz with a bandwidth of 1.4~GHz. This example serves as a means of validation to showcase the feasibility of the structure's functionality and manufacturability. Additionally, this investigation sets the foundation for tackling more intricate scenarios in future works. The resulting frequency response of the infinite FSS is detailed in Fig.~\ref{DesignInfiniteFSS}, revealing transition bands of approximately 100~MHz. The electromagnetic coupling matrix, as well as the infinite FSS response, of the final design are provided in \texttt{MATLAB} format and can be accessed from \cite{EMCouplingMatrixResultsinMATLAB}. Previous iterations in the full-wave optimization loop are detailed in \cite{EMCouplingMatrixResultsinMATLAB}.

As previously discussed, we aim to extract comprehensive physical information from a \emph{single} FEM simulation. Merely obtaining $S$-parameter information from a full-wave simulation may not be sufficient for achieving an optimal design. Therefore, it becomes crucial to understand the actual internal state of the FSS unit cell from an electromagnetic perspective. In our approach, as discussed in Section \ref{sec:EMCouplingMatrix}, we utilize a \emph{single} FEM analysis to derive the electromagnetic coupling matrix, which elucidates the EM behavior among the local EM resonators within the FSS unit cell \cite{delaRubia2022EMCouplingMatrix,delaRubia2022FullWaveCouplingMatrix}. This electromagnetic coupling matrix is then employed for a tailored synthesis of the infinite FSS response directly in the EM domain. Consequently, we obtain a target electromagnetic coupling matrix that serves as our reference to tune the frequency response of the infinite FSS. Our optimization loop is guided by this target electromagnetic coupling matrix, enabling us to achieve the desired electrical response within a \emph{few} iterations, requiring only a tiny number of full-wave FEM simulations.

\subsection{Tolerance Analysis}
The manufacturing stage plays a pivotal role in producing cascade filters, as the potential manufacturing tolerances can significantly influence their performance. A comprehensive tolerance analysis has been conducted to ensure accurate measurement results aligned with the intended design specifications. This analysis aims to assess manufacturing tolerances' impact, guaranteeing the desired performance.

The accuracy restrictions determine that the most convenient fabrication method for such structures is the laser trimming of metal sheets. In this regard, a nominal laser resolution of 25~$\mu m$ is considered, labeled as the variation factor $\Delta s$. Subsequently, an analysis using $2\Delta s$ is conducted to assess the continued validity of the structural performance. The tolerance analysis involves modifying the dimensions of one or several resonators by applying the variation factor $\Delta s$. This approach allows us to examine the evolution of $S_{11}$ from its original design. For each case, the corresponding variable $i$ assumes a value within the range of its nominal value, $s_i$, with a deviation of plus or minus the variation factor $\Delta s$, denoted as $[s_i-~\Delta s, s_i+\Delta s]$.

In our study, we follow a systematic approach. Initially, we analyze the tolerances for individual resonators sequentially, starting from the end and progressing toward the middle (four resonators in total). Later, we examine the two outermost resonators concurrently. Subsequently, we consider the simultaneous analysis of the five central resonators. Finally, we study the variation of all the resonators simultaneously. These specific sets of resonators are chosen to validate or refute the hypothesis that the end resonators exhibit a more significant impact on the degradation of the frequency response caused by manufacturing tolerances.

\begin{figure}[tbp]
    \centering
    \subfloat[]{\includegraphics[width=0.85\columnwidth]{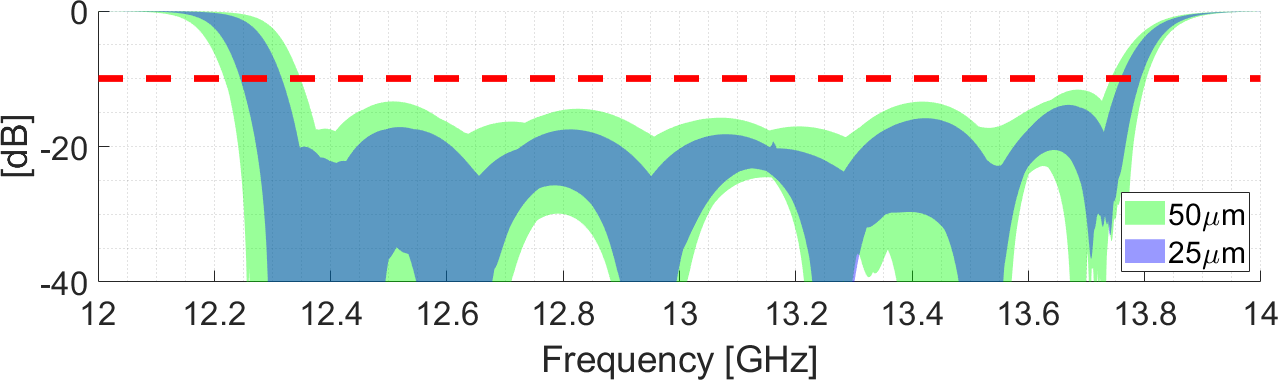} \label{1H}}
    \hfill
    \subfloat[]{\includegraphics[width=0.85\columnwidth]{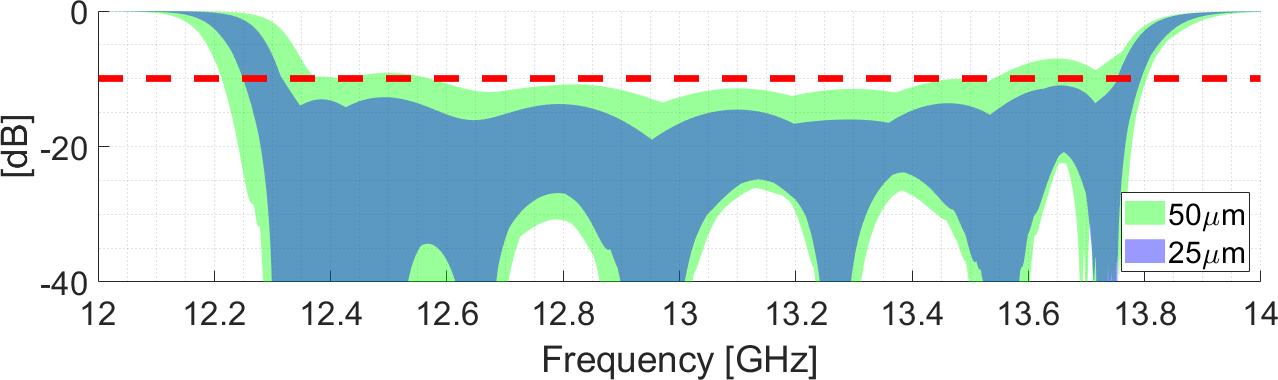} \label{2H}}
    \hfill
    \subfloat[]{\includegraphics[width=0.85\columnwidth]{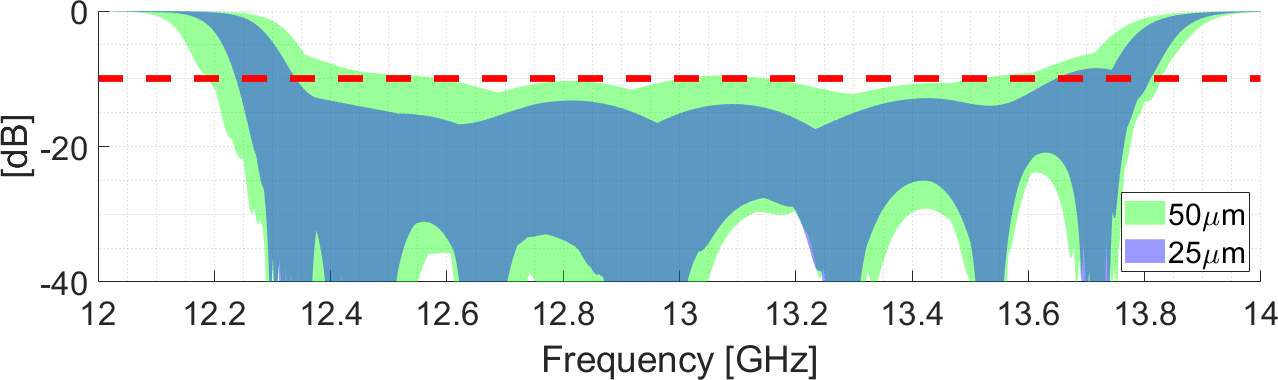} \label{3H}}
    \hfill
    \subfloat[]{\includegraphics[width=0.85\columnwidth]{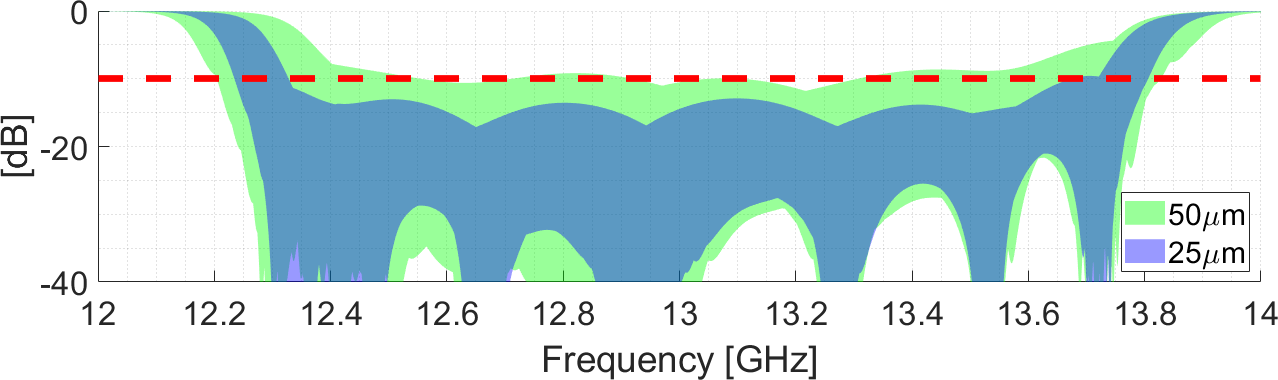} \label{4H}}
    \caption{Tolerance analysis with $\Delta s$ (blue) and $2\Delta s$ (green) for one resonator: (a) first resonator, (b) second resonator, (c) third resonator, and (d) fourth resonator.}
    \label{ToleranciasUnaH}
\end{figure}
%
\begin{figure}[tbp]
    \centering
    \subfloat[]{\includegraphics[width=0.85\columnwidth]{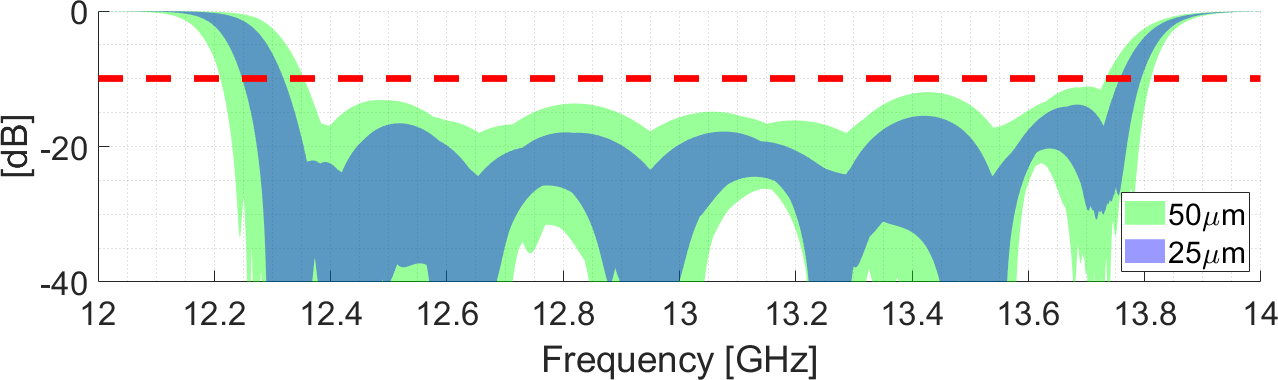} \label{Hextremos}}
    \hfill
    \subfloat[]{\includegraphics[width=0.85\columnwidth]{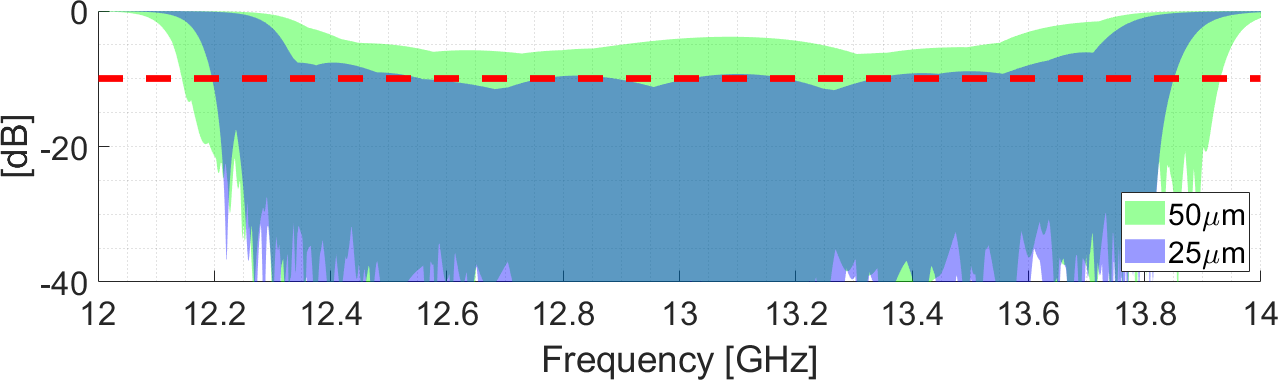} \label{Hcentrales}}
    \hfill
    \subfloat[]{\includegraphics[width=0.85\columnwidth]{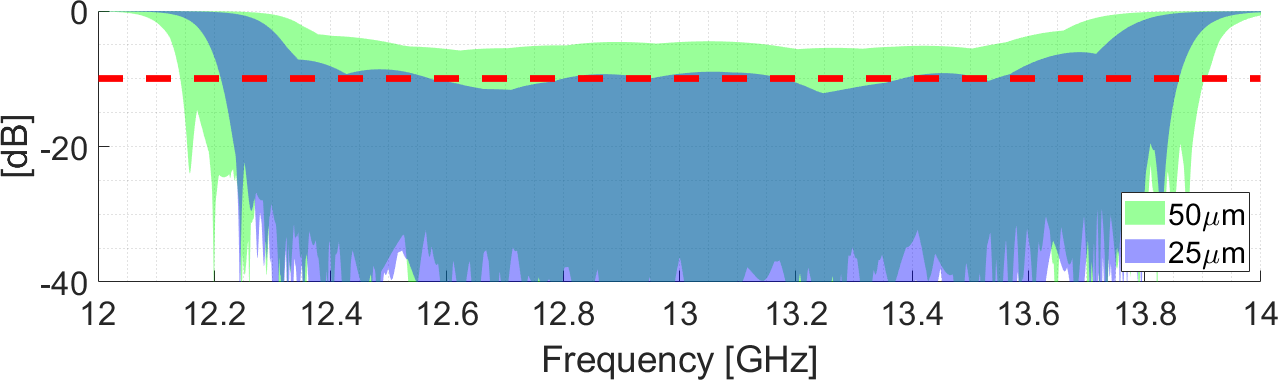} \label{Htodas}}
    \hfill
    \caption{Tolerance analysis with $\Delta s$ (blue) and $2\Delta s$ (green) for multiple resonators: (a) two end-resonators (b) five central-resonators, and (c) all resonators.}
    \label{ToleranciasGruposH}
\end{figure}

Based on these results in Figs. \ref{ToleranciasUnaH} and \ref{ToleranciasGruposH}, it is evident that no specific resonator exhibits a significant effect due to manufacturing tolerances. However, it is observed that the resonators at the ends are comparatively less sensitive. This observation holds significance. Note that the end resonators play a crucial role in achieving optimal performance during the simulation stage. Thus, the design demonstrates robustness against unavoidable manufacturing tolerances.
\section{Experimental Validation} \label{sec:ExperimentalValidations}
\begin{figure}[tbp]
    \centering
    \includegraphics[width=0.90\columnwidth]{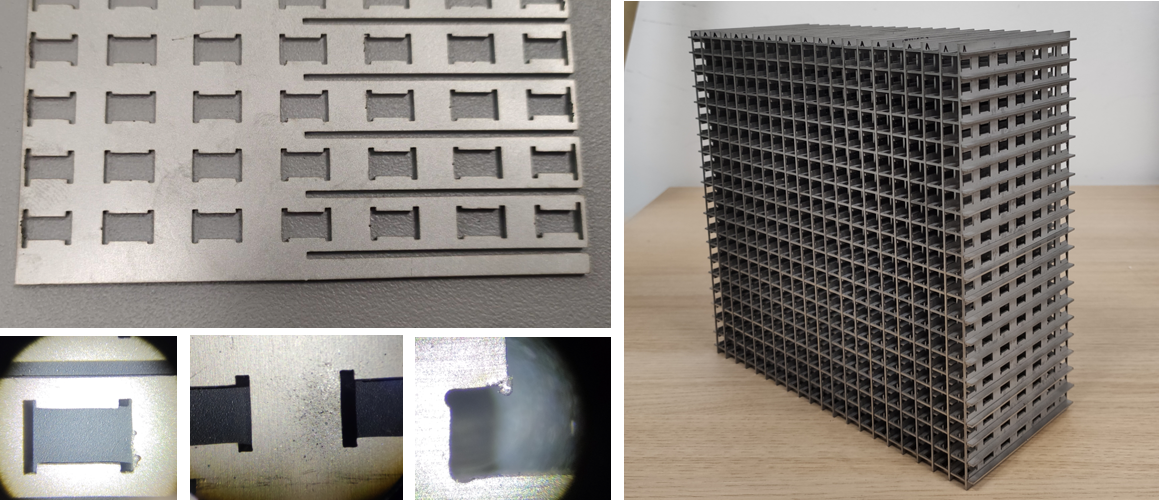} 
    \caption{Details of the metallic plates and the final prototype after the laser trimming process.}
    \label{Placas-laser}
\end{figure}

In order to validate the design results based on the electromagnetic coupling matrix technique, a complete filtering FSS structure is manufactured according to the previous design for the infinite FSS, based on a finite periodic grid of filtering cells made of 7 cascaded resonator cells for two independent polarizations (a 20$\times$20 unit cell prototype is taken into account). For the sake of accuracy, a laser trimming process over steel plates is considered. According to the manufacturer specifications, the trimming tolerances are $\le 30~\mu m$, far below $2\Delta s$. The plates are designed to be assembled as a 3D puzzle, as shown in Fig. \ref{Placas-laser}.
%
\subsection{Experimental set-up and GRL Calibration}
%
\begin{figure}[t!]
    \centering
    \subfloat[]{\includegraphics[width=0.85\columnwidth]{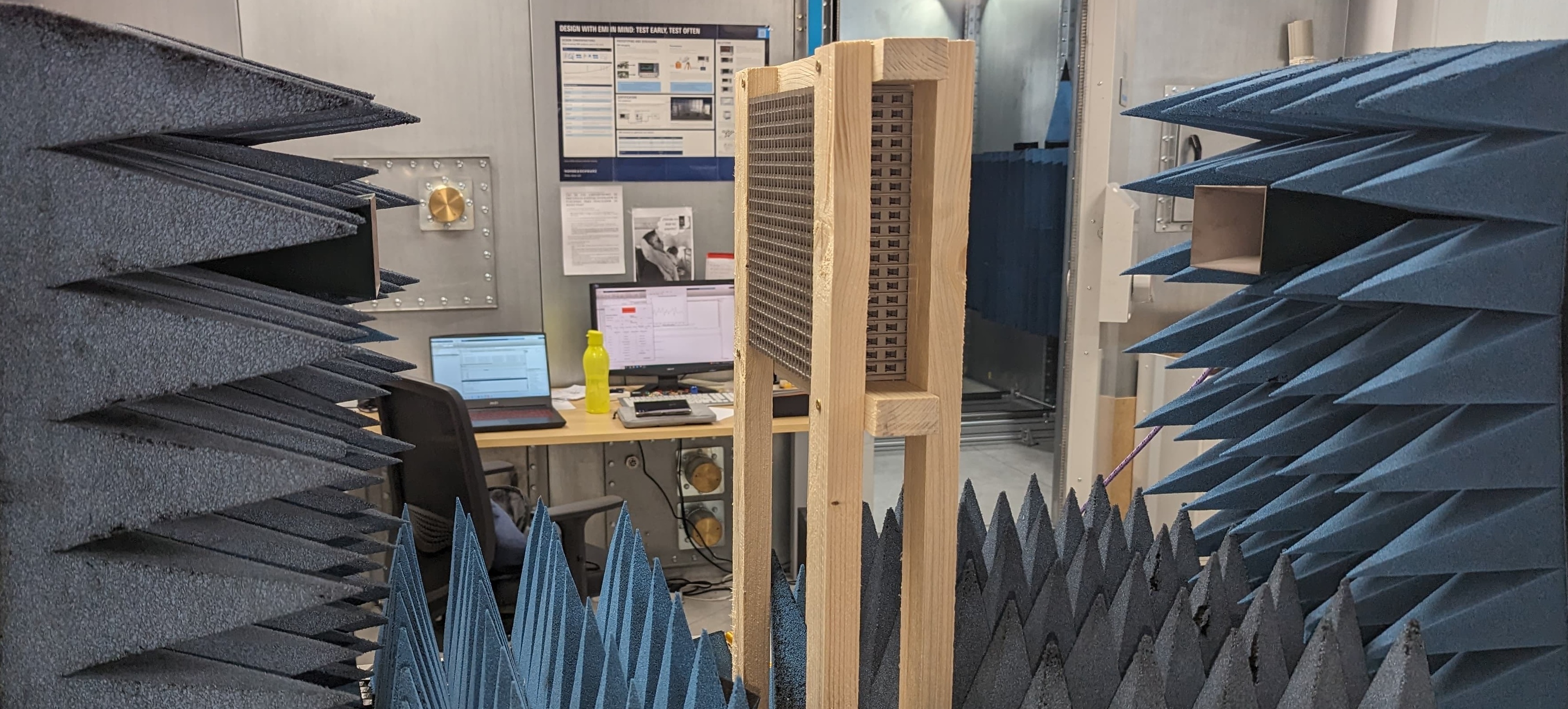} \label{Setup}}
    \hfill
    \subfloat[]{\includegraphics[width=0.85\columnwidth]{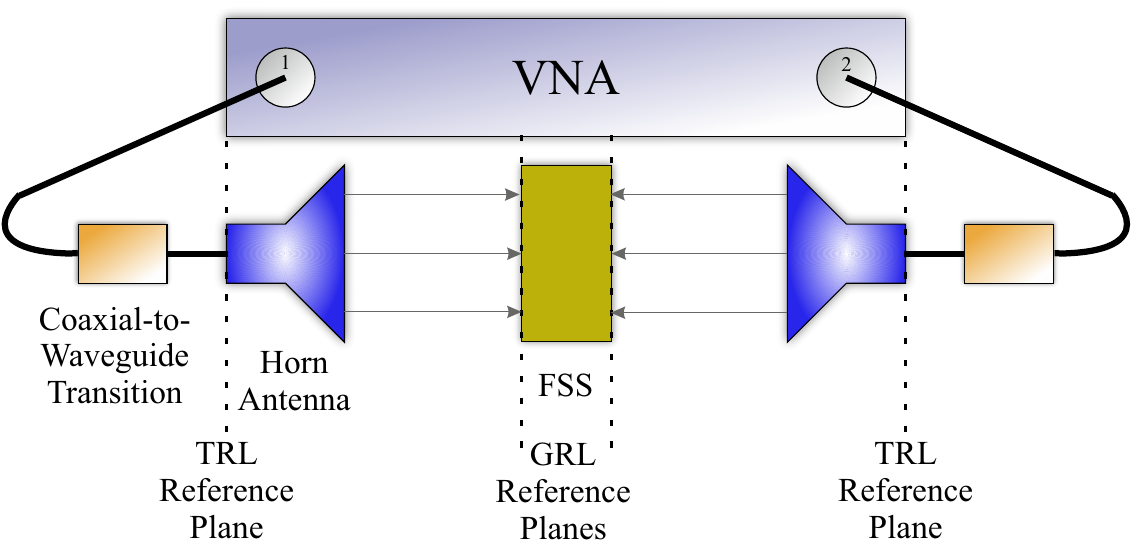} \label{Calibration}}
    \hfill
    \caption{Experimental validation: (a) Smart Wireless
    Technologies Lab (SWT-Lab) set-up, (b) Measuring scheme including the calibration reference planes.}
    \label{Setup-completo}
\end{figure}
The measurements are carried out in the microwave and millimeter measuring facilities of the Smart Wireless Technologies Lab (SWT-Lab) of the University of Granada, as detailed in Fig. \ref{Setup}. Before the characterization process, the set-up must be appropriately aligned and calibrated. Two methods are proposed to be used together to perform the system calibration. First, a TRL calibration is performed at the end of the waveguide feeding the antenna. For this purpose, an aluminum kit has been designed for the WR-75 standard, consisting of a short circuit, which will act as a mirror, and a 2~mm long line. This method places the reference planes at the horn input, specifically at the waveguide-horn transition. After this TRL process, a free space GRL calibration is performed. This calibration, initially proposed in~\cite{Bartley2005} for material characterization, consists of taking two measurements into account: (i) of the empty FSS holder and (ii) of the FSS holder holding a metal plate, in which total reflection of the wavefront is assumed. A similar process to that in the TRL calibration is performed from both measurements. The main difference is that a time-gating process is carried out to isolate the effects of propagation to the sample holder. Within this gating, the effects of the antennas are included. Specifically, a Hamming window is used to include these effects. After calibration, the reference planes are assumed to be in the material on which the normal incidence of a plane wave is occurring. A simplified scheme of the calibration reference planes is shown in Fig.~\ref{Calibration}.
\subsection{Measurement results}
\begin{figure}[tbp]
    \centering
    \includegraphics[width=0.85\columnwidth]{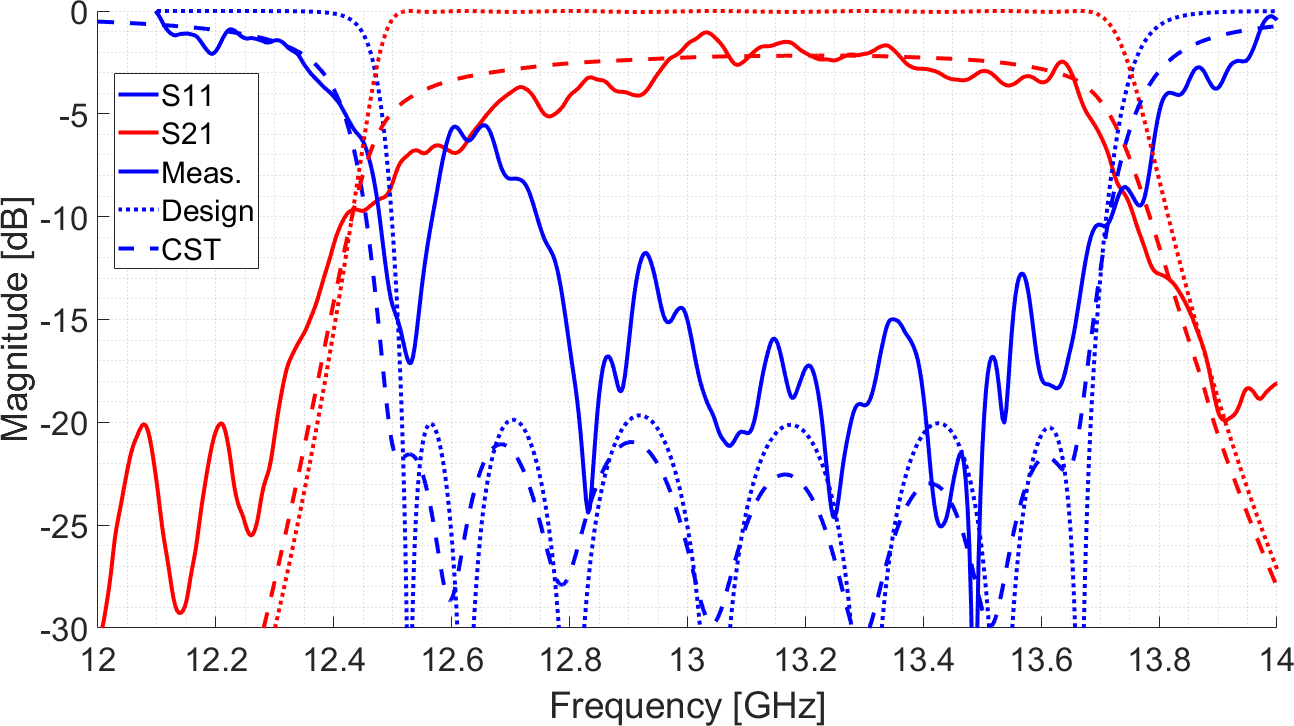}
    \caption{Comparison between design, CST simulation and measurements of $S_{11}$ and $S_{21}$.}
    \label{DesignCSTMeasComparison}
\end{figure}

The filtering structure is characterized through its $S$-parameter matrix behavior measurement, applying the GRL reference plane calibration. The results are shown in Fig.~\ref{DesignCSTMeasComparison}.
%
As it can be seen, 
the experimental results exhibit good frequency agreement with the electromagnetic coupling matrix design when dimension tolerances are considered.
However, it can be seen a light degradation in the structure matching, as well as a slightly rippled transmission behavior. Both effects are justified by the additional mounting tolerances introduced in the 3D mounting of the puzzle structure. The transmission losses are in accordance with the expected ones in a metallic structure, whose roughness and sheet surface defects yield an equivalent conductivity of 0.2·10$^{6}$~S/m.
Upon incorporating these effects into the CST simulation, we observe a better fitting between the simulation results and the experimental measurements.
In any case, these results allow us to validate the electromagnetic matrix coupling method for designing cascaded 3D FSS filtering structures, where manufacturing of the developed prototype has been adequately addressed.
\section{Conclusions} \label{sec:Conclusions}
A wideband dual-polarized fully metallic filter for the Ku-band has been designed and experimentally tested. The design tool is based on a full-wave model of electromagnetic coupling matrix, which allows for the identification of the physical resonators of the filter and the corresponding tuning. The filter under consideration is based on a fully-metallic FSS formed by periodic distributions of square-shaped waveguides with dog-bone-shaped resonators perforated along the walls. The resulting in-line architecture is suitable to control both polarizations in a scenario of high independence. Experimental tests have proven to match the wideband behavior predicted theoretically. Tolerance analysis has been performed, as well as an estimation of the roughness influence in the transmission amplitude to estimate the light degradation exhibited in the passband region.
\bibliographystyle{sty/IEEEtran}
\bibliography{bibliography/references}

\end{document}